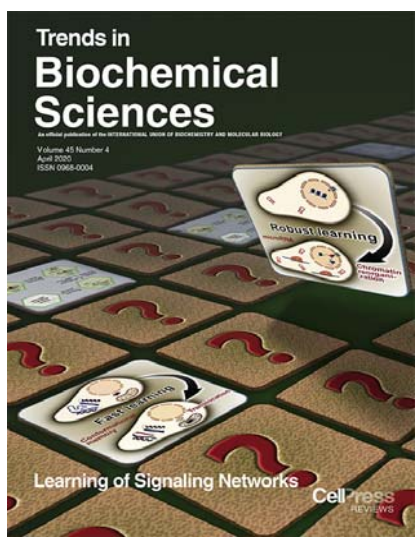

# Learning of signaling networks: molecular mechanisms

Péter Csermely[1*], Nina Kunsic[1], Péter Mendik[1], Márk Kerestély[1], Teodóra Faragó[1], Dániel V. Veres[1,2] and Péter Tompa[3,4]

[1]Semmelweis University, Department of Medical Chemistry, Budapest, Hungary; [2]Turbine Ltd., Budapest, Hungary; [3]Vrije Universiteit, VIB-VUB Center for Structural Biology, Brussels, Belgium; [4]Institute of Enzymology, Hungarian Academy of Sciences Research Centre for Natural Sciences, Budapest, Hungary

**Abstract:** Molecular processes of neuronal learning have been well-described. However, learning mechanisms of non-neuronal cells have not been fully understood at the molecular level. Here, we discuss molecular mechanisms of cellular learning, including conformational memory of intrinsically disordered proteins and prions, signaling cascades, protein translocation, RNAs (microRNA and lncRNA), and chromatin memory. We hypothesize that these processes constitute the learning of signaling networks and correspond to a generalized Hebbian learning process of single, non-neuronal cells, and discuss how cellular learning may open novel directions in drug design and inspire new artificial intelligence methods.

**Highlights**
-- Besides the well-known learning processes of neurons, non-neuronal single cells are able to learn and show a more robust (and often faster) adaptive response when the same stimulus is repeated.
-- Known examples of cellular learning are sensitization- or habituation-type responses.
-- Several molecular mechanisms of neuronal learning, such as conformational memory, protein translocation, signaling cascades, microRNAs, lncRNAs and chromatin memory, also participate in learning of non-neuronal, single cells.
-- We propose that these molecular mechanisms form the integrative memory of signaling networks and display a generalized Hebbian learning process by increasing those edge weights through which the signal has been propagated.

### Learning of non-neural cells: Adaptive Molecular Responses Observed when the same Stimulus is Repeated

Molecular mechanisms of neuronal learning have been well-established in the past decades [1]. However, we know relatively little about the molecular details of learning mechanisms of non-neuronal cells. We define cellular learning as an adaptive response to a simple stimulus observed when the same stimulus is repeated in a short time – as compared to the duration of the cell cycle of the given cell. We note that it is crucial to discriminate between molecular changes which are indeed adaptive, and those which are fortuitous by-products of other, co-occurring adaptive phenomena. To make this distinction, we focus on the molecular mechanisms of adaptation at the unicellular level and do not detail the long-term, multi-step processes of cell reprogramming, development, or disease formation. In these latter cases, many individual learning steps may be interwoven and may reflect adaptation to unrelated events. Due to similar reasons, we do not cover the formation of intergenerational, **epigenetic memory** (see Glossary) and do not detail evolutionary processes. Finally, we focus on non-neuronal cells, since we want to concentrate on the adaptation of the **signaling network** of a single, non-neuronal cell and not that of multi-cellular networks connecting several neuronal cells.

---

[*]Correspondence: csermely.peter@med.semmelweis-univ.hu (P. Csermely)

It is well known from many experimental observations that cellular responses change, i.e. they can become either faster and/or stronger or weaker upon repeated stimuli. For example, budding yeast cells displayed a faster re-activation of the inositol-3-phosphate synthase (Ino1) and galactokinase (Gal1) enzymes after a previous activating stimulus [2] and developed a **molecular memory** of a previous heat stress lasting for several generations [3]. This and a few other examples show that the molecular mechanisms we describe here may go beyond the single cell cycle time-frame of cellular learning as defined in the current paper, and forming a molecular memory of the cell, promote intergenerational, epigenetic learning. We will compare cellular learning and molecular memory formation in the concluding section in detail. In *Arabidopsis*, exposure to the damage signaling hormone, jasmonic acid, caused a stronger response to consecutive dehydration stress [4]. Also in *Arabidopsis*, the molecular memory of a previous heat shock was maintained for several days [5]. Similarly, rice developed molecular memory of drought stress [6]. Additionally, mouse fibroblasts showed a faster and stronger response to the second interferon-β stimulus given one day after the first [7]. Murine $CD8^+$ memory T cells displayed a stronger response if they were re-exposed to *Listeria monocytogenes* 48 hours after the first infection [8]. Importantly, cellular learning can also result in the repression of a response. In budding yeast cells, the *STL1* sugar transporter gene showed a reduced expression to the second hyperosmotic stress as compared to the first [9]. In *Arabidopsis*, a sub-set of *MYC*-dependent genes related to multiple abiotic and hormone response networks did not respond to repeated dehydration stress [10]. Finally, mouse macrophages developed an immune-tolerance after repeated lipopolysaccharide exposure [11]. The activation and repression described in these examples resemble the classical learning types **sensitization** and **habituation** [1], respectively. Regrettably, other **hallmarks of learning**, such as **conditioning** [1], better recognition of the signal from its partial representation, or increased tolerance of noise, have not yet been convincingly demonstrated at the single-cell level.

Here, we provide examples of how different levels of cellular architecture (such as **intrinsically disordered proteins** (**IDPs**), signaling cascades, translocating proteins, RNAs, and chromatin structure) contribute to cellular learning. **Conformational memory** (including that of IDPs), signal integration by signaling cascades and protein translocation may be considered as a faster phase of cellular learning. RNA- (such as microRNA- or long non-coding RNA (lncRNA)-) based molecular memory and many forms of **chromatin memory** develop more slowly but have a longer duration. We demonstrate, through the example of the **epithelial-mesenchymal transition** network [12], how these elements of cellular learning at single cell level are all organized in one signaling network that potentially possesses a learning capability at multiple levels. As the major hypothesis of our paper, we propose that, in signaling networks, cellular learning may be interpreted as a generalized **Hebbian learning** process [1], in which weights of **network edges** of signaling networks where the signal has propagated become increased (i.e. molecular connections become stronger) during the adaptive changes. This novel, integrative understanding of cellular learning may lead to new artificial intelligence and drug design technologies.

**Conformational Memory**
Several proteins display conformational memory, where the protein transiently keeps its active conformation after the dissociation from its former binding partner [13]. Examples include the active state of the endocytosed, unliganded integrin receptor [14], as well as the SERCA Ca-ATP-ase [15]. Here, we propose that conformational memory may participate in the molecular memory formation of single cells (Figure 1). Importantly, the process of increased binding affinity of "protein B" having a conformational memory to its signaling neighbor, "protein A", is the same as the signaling network representation of the Hebbian learning process [1], where the network edge weight of two signal-transducing neighbors (characterizing the strength of their association) will increase because of the signaling process. Several proteins having conformational memory (see below) represent nodes of signaling networks and may have key roles in cellular learning processes.

IDPs are proteins which lack an organized 3-dimensional structure, whereas ID regions (IDRs) are disordered segments (loops, linkers) of at least 20 amino acids in length located in otherwise ordered proteins. Intrinsic disorder can be found in 85% of human signaling proteins [13,16]. Importantly, IDRs regulate organized protein cores in several protein kinases [17], and often act as **molecular switches** that can change the direction of signal propagation [18]. IDRs are enriched in sites of posttranslational modifications and are often alternatively spliced [13], and may have conserved molecular features, such as subcellular localization, membrane transport, motor activity, ribosomal function, etc. [19]. Thus, IDPs/IDRs are good candidates for the conformational memory providing fast cellular learning (Figure 1).

**Prion proteins** are enriched for structural disorder and represent another form of conformational memory. A conformational switch may convert prions to a beta-sheet enriched form making extensive aggregates. Chaperones are required for prion formation but may also erase prion memory in case of severe stress [13]. For example, in budding yeast cells, the prion form of Pin1 maintained the molecular memory of a previous heat stress for subsequent generations [3]. Additionally, the neuronal cytoplasmic polyadenylation element binding (CPEB) protein of the mollusk *Aplysia* can undergo a prion-like conformational transformation and behave as a molecular switch perpetuating molecular memory for years [20]. These observations confirmed the earlier hypothesis that prions may participate in memory formation [21].

**System-level Memory of Signaling Cascades and Protein-protein Interaction Networks**
Francis Crick proposed in 1984 that a signaling complex as simple as a protein kinase and a phosphoprotein phosphatase pair may display molecular memory, preserving its active or inactive state despite the turnover of its constituent proteins [22]. Later studies defined a prominent role of molecular switches in molecular memory formation (including the role of IDPs) [18]. A recent model uncovered how larger segments of the signaling network develop cooperation-based, **system-level memory**. The MAPK cascade displays a rich repertoire of transient adaptive responses characterized by both frequency and amplitude modulations. Different relaxation rates of cascade components lead to 'post-activation bursts' keeping the cascade in an 'activation-competent' state. This can form a system-level memory of the first activation making later responses faster and more robust [23]. Such a short-term molecular memory was demonstrated in the yeast osmotic stress response, too. If osmotic stress was repeated within several minutes, members of the Hog1 signaling pathway were still phosphorylated and thus 'awaited' the next signal in a pre-activated state [24]. These examples show how the concerted activation of signaling cascades may contribute to cellular learning.

Protein-protein interactions constitute an essential element of signaling networks. Yet, many of them are not directly involved in building up signaling networks, but rather function in modifying their behavior. Weak protein-protein interactions give rise to 'noise' that diminishes the efficiency of information transmission. Increased interaction strength helps information transmission, but slows down response dynamics showing a trade-off between efficiency and responsiveness [25]. Molecular chaperones increase the frequency of out-of-equilibrium states and help the 'disorganization' of protein segments [26]. Thus, chaperones may act both as facilitators of molecular memory formation and as a 'forgetting mechanism'. These examples show how protein-protein interactions may fine-tune cellular learning of signaling networks.

**Subcellular Protein Translocation**
Signal-induced **protein translocation** (triggered by, e.g., phosphorylation) between two cellular compartments is a widespread phenomenon potentially affecting thousands of human proteins [27]. Protein translocation is actively involved in the reconfiguration of signaling networks in cellular learning processes. For example, inhibition of NF-κB p65 nuclear translocation disrupted the formation of both $CD8^+$ memory T cells and memory B cells [28,29]. Further, protein kinase C βII-induced upregulation and mitochondrial translocation of the adaptor protein, p66SHC, was associated with the formation of hyperglycemic molecular memory of human aortic endothelial cells (Figure 2 [30]). Protein translocation may also occur between subcompartments of a cellular organelle, such as in the nucleus or in the form of the formation of biomolecular condensates by liquid-liquid phase separation [31]. Protein translocation establishes a whole set of novel protein-protein interactions, increasing their edge weights in signaling networks.

**Moonlighting proteins** perform multiple functions, often in different locations, resulting from protein translocation [27]. For example, multiple interactions between the moonlighting immunomodulatory activities of acute phase proteins and monocyte-derived dendritic cells play a key role in forming immunological memory [32].

**RNA-based Molecular Memories**
Various types of RNAs were also shown to participate in cellular learning processes. Since RNAs have a short lifetime, their *de novo* transcription is needed to initiate their effects. MicroRNAs decrease the protein expression noise of hundreds of lowly expressed proteins [33], increasing the noise-tolerance and thus robustness of cellular learning by reducing gene expression variability. Molecular memory formation was shown by microRNA-156 which posttranscriptionally down-regulated SPL transcription factor genes in the plant *A. thaliana*, causing the development of thermotolerance and thus conferring a molecular memory of a previous heat shock. This molecular memory was maintained for several days (Figure 3A) [5]. Additionally, microRNA-221 and -222 induced inhibition of macrophage activity during the development of lipopolysaccharide tolerance [11]. The microRNA cluster 17-92 was transiently induced after T cell activation. Both the induction and later silencing of the microRNA 17-19-cluster were mandatory to the development of $CD8^+$ memory T cells [34]. Further, microRNA-21 was involved in the development of fibrotic mechanical memory of mesenchymal stem cells [35]. These examples show the widespread involvement of microRNAs in both sensitization- and habituation-type cellular learning processes.

MicroRNA induction can be perceived in signaling networks as an increased edge weight of participating microRNAs.

Other types of RNA have been shown to contribute to cellular learning. In particular, lncRNAs participated in forming molecular memory of rice drought stress [6] and the development of $CD8^+$ memory T cells after viral infection [36]. In contrast, a lncRNA originating at -2700 upstream of the budding yeast HO endonuclease erased previous molecular memory of nutrient deprivation- or pheromone-induced cell cycle arrest [37]. These examples elucidate the RNA-dependent regulation of molecular memory formation, showing the richness of the contribution of various RNAs to cellular learning processes in signaling networks.

**Chromatin Memory**
Histone modification (including histone-methylation, -phosphorylation, -acetylation, -ubiquitylation, and -sumoylation), as well as DNA methylation (occurring at adenine and cytosine nucleotides and often forming CpG dinucleotides, especially in mammals) also play key roles in cellular learning. These processes are also called **transcriptional memory**. Changes in histone acetylation can occur on a much faster time scale than those in DNA methylation [38]. Lysine methylation of histone H3 participates in both sensitization and habituation of *Arabidopsis* [4,10], in sensitization of mouse fibroblasts and human HeLa cells by interferon-β and -γ, respectively [7,39], as well as in $CD8^+$ memory T cell formation [8]. In the study of Komori *et al* [40], 466 CpG dinucleotides of 132 genes displayed differential DNA methylation between naive and memory $CD4^+$ T lymphocytes. Erasure of DNA methylation ('forgetting') can be performed *via* ten-eleven translocation (TET) DNA demethylases [41].

The 3-dimensional chromatin structure also plays an important role in cellular learning. Sensitization to hyperosmotic stress was abrogated if the reporter gene was placed to a pericentromic chromatin domain in yeast cells [9]. Nup2-mediated association of the *INO1* and *GAL1* genes with the nuclear pore complex and histone modifications led to the rapid re-activation of *INO1* and *GAL1* genes after a repeated signal. Both the Set1/COMPASS methyltransferase complex and the Mediator complex were remodelled in these processes (Figure 3B) [2]. The human major histocompatibility complex class II gene DRA was persistently relocated to promyelocytic leukemia nuclear bodies after interferon-γ treatment causing a sensitization to a subsequent interferon-γ stimulus [39]. Increased transcription by changes in chromatin organization can be perceived as increasing edge weights of transcription factors in signaling networks. It is a question of future studies, whether shape fluctuations [42] or rotation of cell nucleus [43] also play a role in cellular learning processes.

**Learning of signaling networks**
All the processes we described so far constitute changes of signaling networks playing a role in integrative cellular learning and molecular memory formation. We illustrate the potential cellular learning mechanisms of signaling networks by the epithelial-mesenchymal transition network of hepatocellular carcinoma cells as described by Reka Albert and her group (Figure 4, Key Figure) [12]. As shown in Figure 4, many nodes of this signaling network may participate in one or more mechanisms of cellular learning (i.e. possessing intrinsic disorder, participating in translocation, being an RNA, or being a protein regulated by chromatin changes). Note that these adaptive changes all re-calibrate the edge weights of signaling networks. Increasing the edge weights of those connections of the signaling network, which have been activated by the incoming signal, corresponds to a Hebbian learning process of the signaling network of single, non-neuronal cells. We note that e.g. in case of a habituation response certain edge weights may also decrease, which is a molecular example of **anti-Hebbian learning**. The demonstration of learning of signaling networks and the extension of the Hebbian learning process to the molecular level of single cells are the major novel concepts of our paper.

Several observations showed that epithelial-mesenchymal transition has a molecular memory [44-46]. Purple arrows in Figure 4 point to those nodes, which have already been identified as participants in these processes. Importantly, all these nodes possess one or more features that pertain to the potential mechanisms of cellular learning discussed herein.

Signaling networks may be extended by cytoskeletal [47] and inter-organelle [48] networks, as well as by inter-cellular signaling [49], filamental [50], and membrane [51] networks. These networks may all have a potential, heretofore not exactly described, role in promoting cellular learning of non-neuronal cells.

**Concluding Remarks**
In this paper, we showed how conformational memory of proteins, signaling cascades, subcellular protein translocation, various RNA molecules, and chromatin memory can result in integrative learning of signaling networks in single, non-neuronal cells. We hypothesize that signaling networks of non-neuronal cells display features of Hebbian learning [1] by increasing the strength of molecular connections between signaling molecules involved. We believe the examples outlined herein demonstrate that various molecular mechanisms develop two major types of cellular learning: sensitization and habituation. However, the direct demonstration of more complex forms of cellular learning remained notoriously difficult in non-neuronal cells.

In a network description, a neuron corresponds to a single node at the neuronal network level, while the same neuron contains, as one of its segments, its own signaling network. (Thus in the pre- and postsynaptic neurons, two of these signaling networks become joined.) We would like to emphasize that learning at both levels, the single-cell signaling network and the multi-cellular neuronal network, uses the same underlying molecular mechanisms elucidated here (such as conformational memory, signaling cascades, protein translocation, microRNA, and chromatin memory). However, neuronal learning displays several molecular forms even in a single neuron (such as synaptic densities, changes of membrane potentials, etc.) which are not characteristic of the cellular learning process of a single, non-neuronal cell. Obviously, neuronal learning also mobilizes the enormous potential of multi-cellular, neuronal networks, which, by definition, can not be reached at the single-cell level. Thus evidently, multi-cellular, neuronal networks allow the development of incomparably more sophisticated learning processes than those of single, non-neuronal cells described in this paper.

The formation of transgenerational (epigenetic) memory also uses many of the molecular mechanisms of the intra-generational, cellular learning listed here (e.g. DNA and histone methylation, and related chromatin rearrangements [2,40,41,52], as well as protein compartmentalization [53], microRNAs [54], and prions [3]). These mechanisms build up the molecular memory of the individual cell lasting for single or multiple cell cycles. However, short-term changes, such as conformational memory of IDPs and changes of signaling cascades, may not be extended for multiple cell cycles and thus may only participate in the *sensu strictu* cellular learning we defined in this paper and not in epigenetic memory formation. We note that later experiments will certainly provide a solid basis to extend the molecular mechanisms of cellular learning much beyond a single cell cycle.

It is an open question (see Outstanding Questions) whether different types of cellular learning (e.g. sensitization and habituation) proceed via different or similar mechanisms. Current data have not yet been examined in detail for elucidating the molecular mechanism(s) of these phenomena in the same system. Another important open question is how forgetting of cellular learning proceeds. Forgetting introduces the **Oja's rule** to Hebbian learning preventing the 'over-excitation' of the network due to the continuous growth of its edge weights.

A better understanding of cellular learning processes will inspire progress in several areas of science. A recent paper on non-Markovian chemical reaction networks on gene expression showed that molecular memory of protein synthesis and degradation may induce feedback, bimodality, switch-behavior and may fine-tune gene expression noise [55]. These findings open the possibility that our concept of generalized Hebbian learning may be extended to metabolic and other types of molecular networks in the future. Chemically induced proximity between two adjacent signaling molecules by a drug became a recent drug design paradigm [56]. Enhanced proximity in these therapeutic approaches may also mimic the effect of cellular learning. Chromatin modifier drugs are already used in anti-cancer therapy [57]. We expect a much wider use of drugs targeting the cellular learning mechanisms described in this paper in the future. In the analogy of genetic algorithms and neural networks, cellular learning may also inspire novel artificial intelligence methods. Cellular learning is a research area which will show dramatic progress in the coming years, and we are happy to invite our colleagues to join these efforts.

---

**Outstanding questions**
-- Are there different types of molecular mechanisms for different types of cellular learning, like sensitization or habituation?
-- How are the elements of the Hebbian learning process of cellular learning, highlighted in this paper, coordinated at the signaling network level?
-- Are the mechanisms of cellular forgetting (i.e. erasure of molecular memory) coordinated at the signaling network level?
-- What is conformational memory's contribution to cellular learning?
-- How many intrinsically disordered regions do actually fold after a signal-dependent binding event to the signaling partner protein or to the membrane?
-- How long does the folded intrinsically disordered region stay folded after the signal's termination, thus (presumably) the dissociation of the signal-induced protein complex?
-- What is the role of cytoskeletal, inter-organelle and inter-cellular networks as well as liquid-liquid phase separation in the formation of molecular memory?
-- How does nuclear dynamics contribute to cellular learning?

---

**Glossary**
-- **Anti-Hebbian learning**: a learning process where edge weights are not increasing (as in Hebbian learning) but decreasing.
-- **Chromatin memory:** altered 3-dimensional structure of the chromatin which provides different accessibility of genes for transcription after a repeated signal; histone and DNA modifications may play a role in its development.
-- **Conditioning:** a learning procedure in which stimulus A is paired with stimulus B, and the response to stimulus B becomes activated after stimulus A alone.

-- **Conformational memory:** an 'active' (e.g. high-affinity, binding-competent) conformation of a protein which is adopted upon binding to its partner (which may be a substrate, another protein, or a membrane) and does not relax back to the original, 'inactive' conformation before the next activation of the same protein.
-- **Epigenetic memory:** a heritable, intergenerational change in gene expression or behavior that is induced by a previous signal.
-- **Epithelial-mesenchymal transition:** a process by which epithelial cells lose their cell polarity and cell-cell adhesion, and gain migratory and invasive properties to become mesenchymal cells; occurring (among others) in wound healing, organ fibrosis and initiation of metastasis in cancer progression.
-- **Hallmarks of learning:** sensitization, habituation, or conditioning are the major hallmarks of learning processes; in addition, discrimination of different signals, reconstruction of the signal from its partial representation, and noise-tolerance are also considered as additional hallmarks.
-- **Habituation:** a form of non-associative learning where a response to a stimulus decreases after repeated or prolonged presentations of that stimulus.
-- **Hebbian learning:** in the current paper the classical, neuronal Hebbian learning is generalized as an intracellular process of single, non-neuronal cells, which increases those edge weights (i.e. strengths of molecular connections) of signaling networks, where the signal has propagated.
-- **Intrinsically disordered proteins (IDPs):** proteins that do not have an ordered 3-dimensional structure; in many proteins, structural disorder only extends to a segment (intrinsically disordered region, IDR) of the protein.
-- **Molecular memory:** molecular mechanisms inducing cellular learning either within a single cell cycle or by developing intergenerational, epigenetic memory.
-- **Molecular switch:** a molecule that can be reversibly shifted between two or more stable states.
-- **Moonlighting proteins:** proteins which perform multiple functions often in different cellular locations and/or participate in different protein complexes.
-- **Network edge:** connection between two network nodes (i.e. basic elements of the network); network edges may be weighted, directed, and may have sign (i.e. they may encode activation or inhibition).
-- **Oja's rule:** introduces a 'forgetting' term to the Hebbian learning rule making sure that the sum of total edge weights should not increase.
-- **Prion proteins:** misfolded proteins able to transmit their misfolded conformation to normal variants of the same protein.
-- **Protein translocation:** signal-induced re-localization of proteins between subcellular compartments.
-- **Sensitization:** a form of non-associative learning where a repeated stimulus results in the amplification of its response.
-- **Signaling network:** a directed network of proteins and RNAs participating in cellular signaling processes.
-- **System-level memory:** a form of molecular memory which is not provided by individual signaling molecules, but by the concerted activation of signaling pathways.
-- **Transcriptional memory:** a set of modifications of DNA and DNA-binding proteins (primarily: histones) regulating the accessibility of genes for transcription.

**Acknowledgments**
We thank the members of the LINK network science group (http://linkgroup.hu) for their helpful comments, Jason H. Brickner (Northwestern University, Evanston IL, USA) for his help to summarize the mechanism shown in Figure 3B and the anonymous reviewers for their excellent suggestions to better our manuscript. This work was supported by the Hungarian National Research, Development and Innovation Office [K115378 and K131458 PC; K124670 PT], by the Higher Education Institutional Excellence Programme of the Ministry of Human Capacities in Hungary, within the framework of the Molecular Biology thematic programmes of the Semmelweis University, by Artificial Intelligence Research Field Excellence Programme of the National Research, Development and Innovation Office of the Ministry of Innovation and Technology in Hungary (TKP/ITM/NKFIH), by the Odysseus grant G.0029.12 from Research Foundation Flanders (FWO), a "Korea-Hungary & Pan EU consortium for investigation of IDP structure and function" from National Research Council of Science and Technology (NST) of Korea [NTM223161 PT].

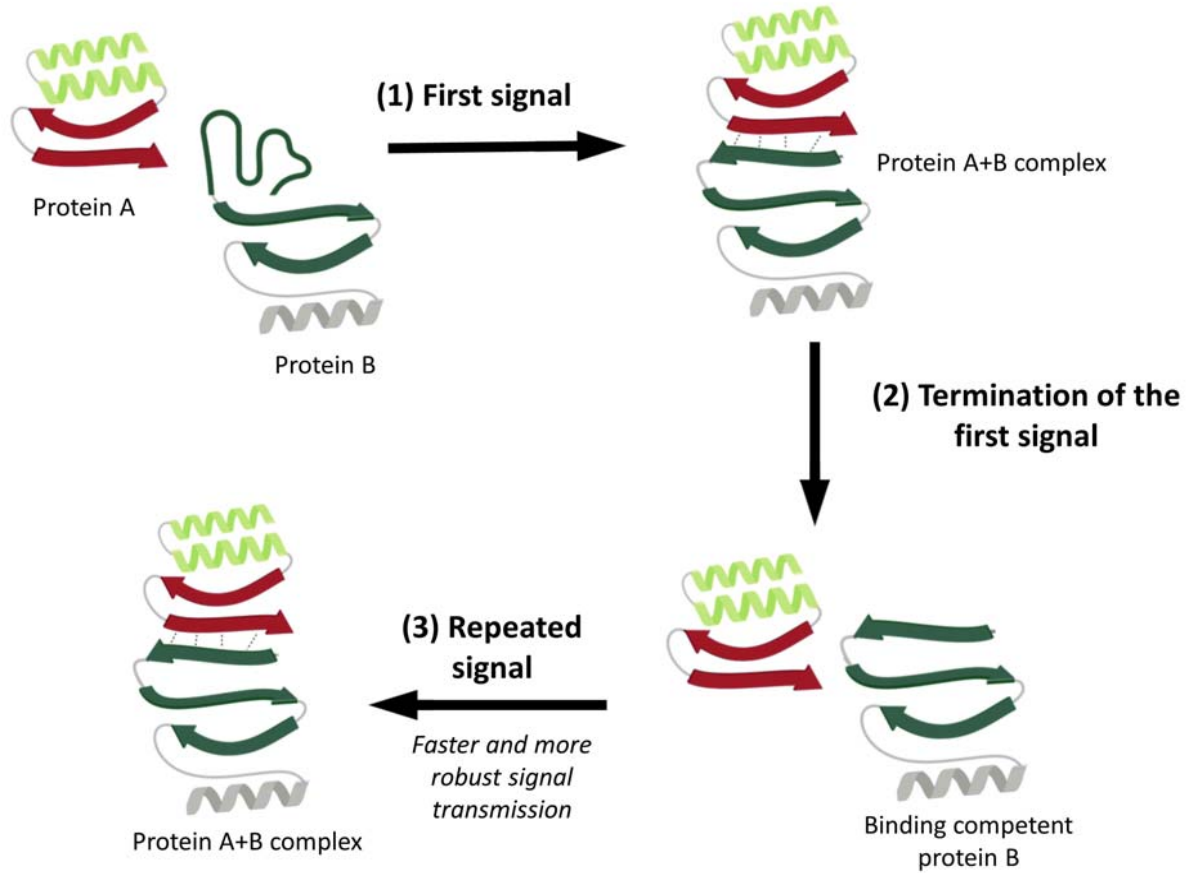

**Figure 1. Conformational memory of signaling proteins as a potential form of cellular learning.** Conformational memory, where proteins transiently keep their binding competent state after dissociation, is a well-established phenomenon [13-15]. For example, the integrin receptor (β1 subunit) [14], SERCA Ca-ATP-ase [15], and prion-like proteins [3,20,21] all possess conformational memory and participate in cellular learning. Here, we illustrate the steps of a conformational memory-mediated learning process on the example of an intrinsically disordered protein (IDP). Importantly, 85% of human signaling proteins contain intrinsically disordered regions (IDRs), which opens the possibility for the transient stabilization of their signal-induced folding [13,16]. (1) The first signal induces the association of neighboring proteins A and B in a signaling cascade which induces a binding-competent conformation of protein B (e.g. via folding of an IDR of protein B) [58]. (2) After the first signal's termination, proteins A and B dissociate. However, within a time window (which may depend on the unfolding rate of the IDR of Protein B), protein B keeps its binding-competent conformation as a conformational memory. (3) If the first signal is repeated soon, the second signal finds protein B still in a binding-competent state, which causes a faster and more robust signal transmission. The signal-induced conformational memory of protein B increases the binding affinity between protein A and protein B. Note, this is exactly the same as the signaling network representation of the Hebbian learning rule [1], where the edge weight of two signal transducing neighbors increases because of the signaling process.

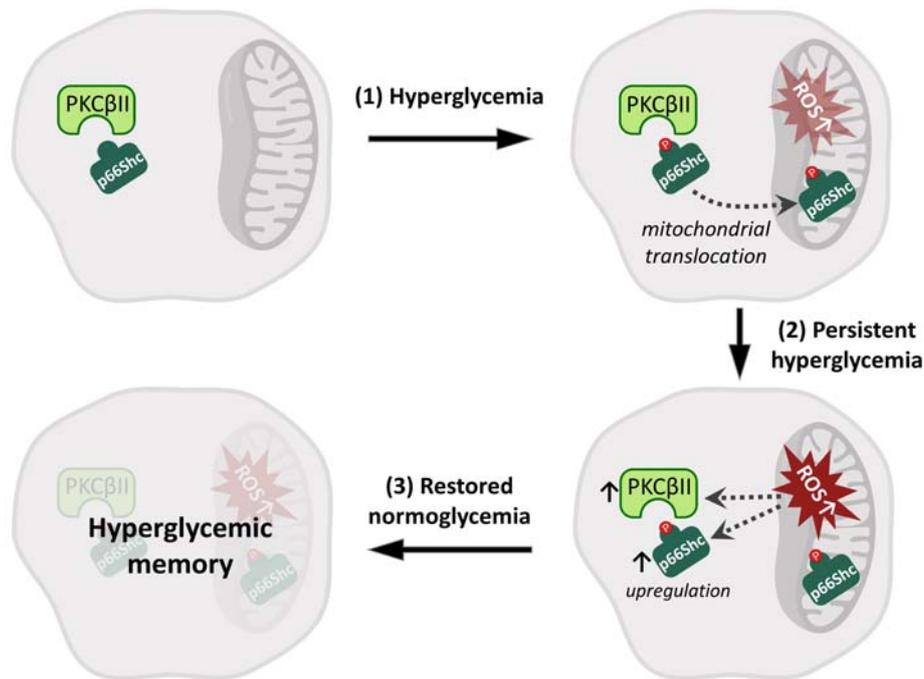

**Figure 2. Subcellular translocation as a form of cellular learning**. Signal induced translocation of proteins between subcellular compartments is a wide-spread phenomenon [27]. Many of these processes may participate in cellular learning. As an example, here we show the protein kinase C βII (PKCβII)-induced upregulation and mitochondrial translocation of the adaptor and reactive oxygen species (ROS) sensor protein, p66Shc, which is associated with forming hyperglycemic molecular memory of human aortic endothelial cells [30]). (1) Hyperglycemia leads to PKCβII-induced phosphorylation and mitochondrial translocation of p66Shc, which induces ROS production. (2) Persistent hyperglycemia upregulates ROS and, consequently, PKCβII and translocated p66Shc, which leads to a vicious circle. (3) After restored normoglycemia p66Shc remains in the mitochondria causing a hyperglycemic molecular memory characterized by increased production of ROS.

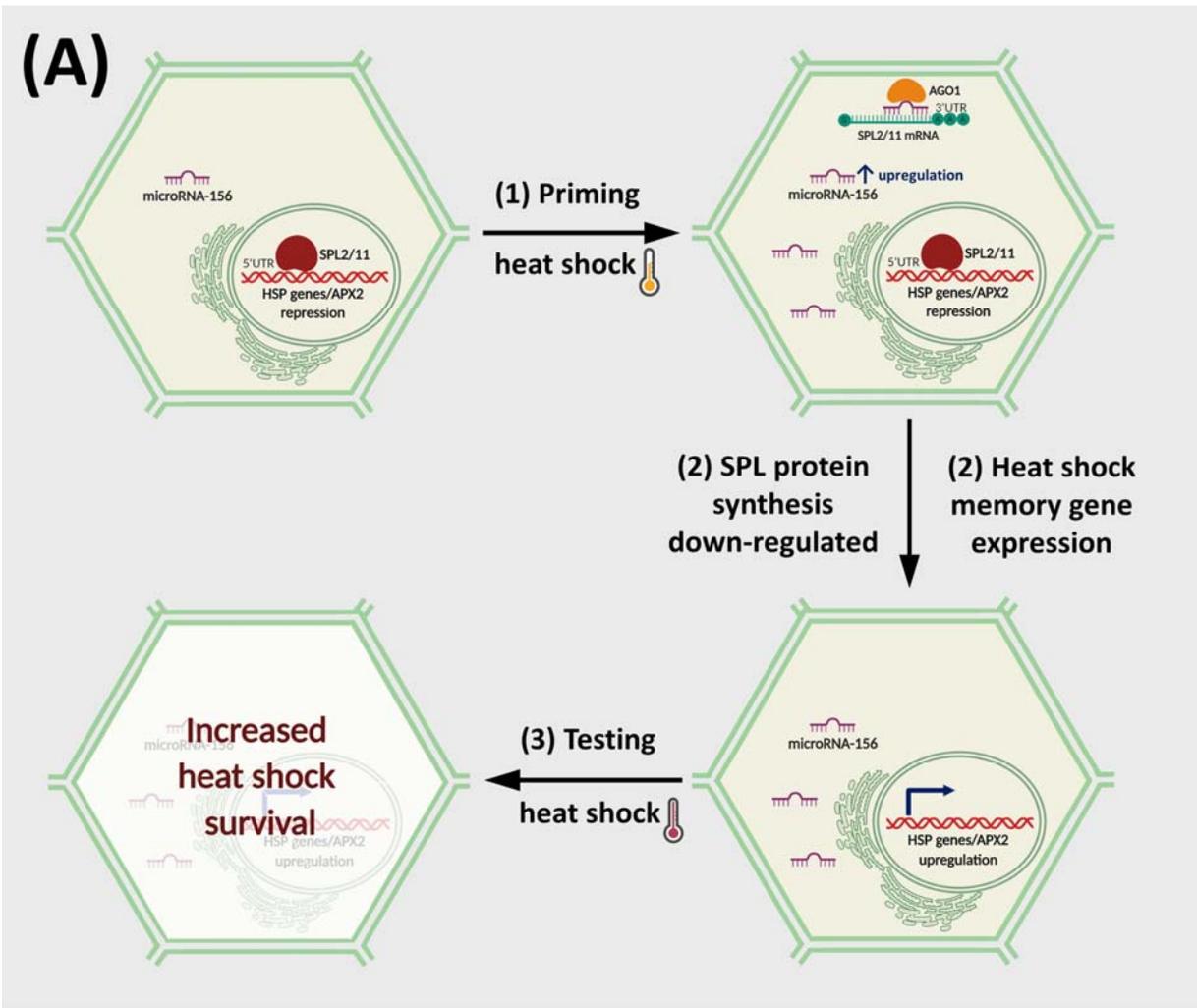
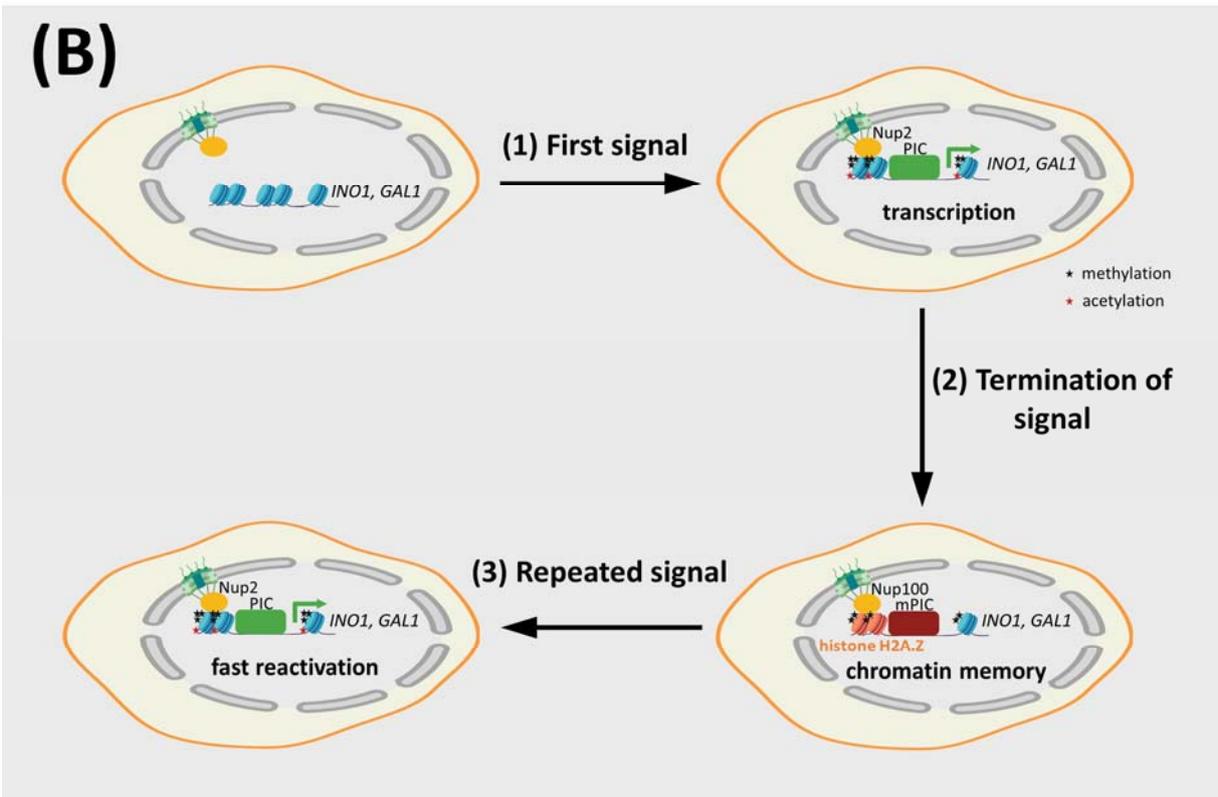

**Figure 3. Various forms of RNA-based and chromatin-memory**. (A) RNA-based molecular memory. Many types of RNA molecules (such as microRNA and lncRNA) participate in cellular learning. As an example, (1) the priming heat shock upregulated microRNA-156 (with the help of the argonaute RNA-induced silencing complex, AGO1), which (2) posttranscriptionally down-regulated the SPL2/11 transcriptional repressor, allowed the synthesis of heat shock proteins (HSP) and ascorbate peroxidase (APX2) in the plant *Arabidopsis thaliana*. These molecular mechanisms caused the development of thermotolerance, thus (3) induced an increased survival upon a second, larger, lethal heat shock due to the previous, priming heat shock. This molecular memory was maintained for several days [5]. (B) Chromatin memory. Chromatin segments remain open and accessible sustained long after the repeated signal due to persistent histone acetylation and DNA de-methylation [8]. Gene-demarcation and gene-association with the nuclear pore complex are forms of global chromatin rearrangements leading to the development of molecular memory. As an example, (1) the yeast genes of inositol-3-phosphate synthase (*INO1*) and galactokinase (*GAL1*) associate with the nuclear periphery via the nuclear pore complex component, Nup2. (2) This association together with histone methylation, acetylation, the incorporation of the specific histone variant, H2A.Z, as well as a modified preinitiation complex (mPIC) lacking the Kin28 CTD kinase, (3) led to a rapid re-activation of *INO1* and *GAL1* after a repeated signal [2]. Panel B was extended and adapted using Figure 8 of reference [2] with permission.

**Key Figure**
Learning of signaling networks

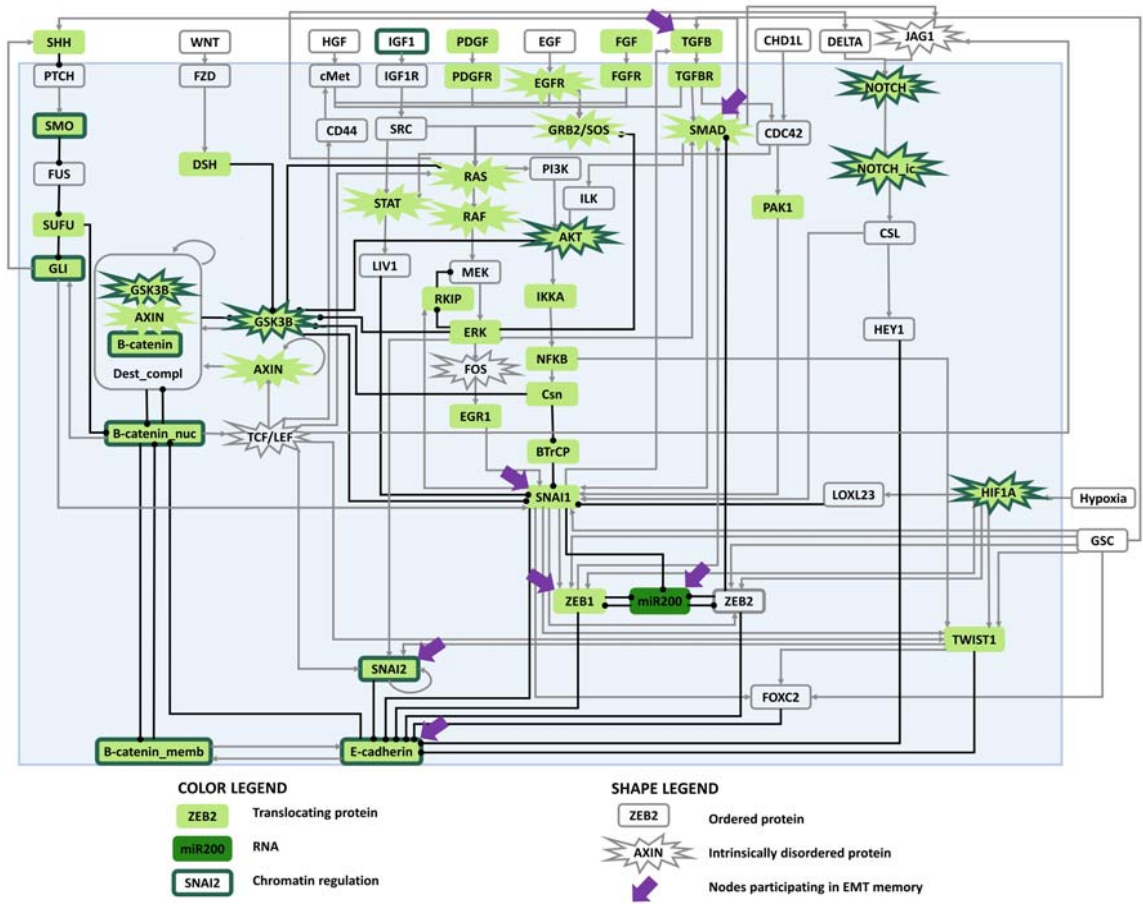

**Figure 4, Key Figure.** Learning of signaling networks. We illustrate various potential cellular learning mechanisms in the signaling network regulating epithelial-mesenchymal transition of hepatocellular carcinoma cells [12]. The large pale blue rectangle represents the hepatocellular carcinoma cell. Grey arrows and black dot-head arrows mark activations and inhibitions, respectively. Intrinsically disordered proteins (identified using the DisProt database [59]), which may possess conformational memory, are marked with starbursts. Proteins potentially participating in subcellular translocation (identified as high confidence translocating proteins in the Translocatome database [27]) are marked with light green rectangles. The participating microRNA is marked with a dark green rectangle. Targets of chromatin regulators in hepatocellular carcinoma (collected from the CR2Cancer database [60]) are marked with dark green edges. Note that epithelial-mesenchymal transition has many more participating RNAs [61] than microRNA-200 of the original network [12]. The addition of more RNAs and chromatin regulators (like histone modifiers or DNA methylases) will be a logical step of future work. Though the examples depicted are not complete, it is obvious that many nodes of this signaling network may participate in one or more mechanisms of cellular learning. Purple arrows highlight those nodes, which have already been identified as part of the mechanisms inducing epithelial-mesenchymal transition memory [44-46]. All of these nodes possess one or more features identified as potential mechanisms of cellular learning in our paper.